\documentclass[twocolumn,prd,aps,showpacs,preprintnumbers,nofootinbib,eqsecnum]{revtex4}

\usepackage{amsmath,amsfonts,amssymb}
\usepackage{wrapfig}           % gopu
\usepackage{graphicx}          % Include figure files
\usepackage{color}             % gopu
\usepackage{dcolumn}           % gopu
\usepackage{psfrag}
\usepackage{pstricks,pst-plot} % for colors like {\blue ...}

\usepackage{t1enc}
\usepackage[english]{babel}   
\usepackage[latin1]{inputenc}

\allowdisplaybreaks

\newcommand{\be}{\begin{equation}}
\newcommand{\ee}{\end{equation}}
\newcommand{\bea}{\begin{eqnarray}}
\newcommand{\eea}{\end{eqnarray}}

\newcommand{\ben}{\begin{eqnarray*}}
\newcommand{\een}{\end{eqnarray*}}

\newcommand{\bs}{\begin{subequations}}
\newcommand{\es}{\end{subequations}}

\newcommand{\vek}[1]{\boldsymbol{#1}}

\begin{document}

%\preprint{astro-ph/(2005)}

\title{Parametric derivation of the observable
relativistic periastron advance for binary pulsars}

\author{Christian K\"onigsd\"orffer}
\email{C.Koenigsdoerffer@uni-jena.de}
\affiliation{Theoretisch-Physikalisches Institut,
Friedrich-Schiller-Universit\"at Jena, Max-Wien-Platz 1,
07743 Jena, Germany}

\author{Achamveedu Gopakumar}
\email{A.Gopakumar@uni-jena.de}
\affiliation{Theoretisch-Physikalisches Institut,
Friedrich-Schiller-Universit\"at Jena, Max-Wien-Platz 1,
07743 Jena, Germany}

\date{\today}

\begin{abstract}
We compute the dimensionless relativistic 
periastron advance parameter $k$, which is measurable
from the timing of relativistic binary pulsars.
We employ for the computation the recently derived
Keplerian-type parametric solution
to the post-Newtonian (PN) accurate conservative dynamics
of spinning compact binaries moving in eccentric orbits.
The parametric solution and hence the parameter $k$ are
applicable for the cases of \emph{simple precession}, namely,
case (i), the binary consists of equal mass compact objects,
having two arbitrary spins, and
case (ii), the binary consists of compact objects of arbitrary mass,
where only one of them is spinning with an arbitrary spin.
Our expression, for the cases considered,
is in agreement with a more general formula 
for the 2PN accurate $k$,
relevant for the relativistic double pulsar PSR J0737--3039,
derived by Damour and Sch\"afer many years ago,
using a different procedure.
\end{abstract}

% 04.25.Nx: post-Newtonian
% 97.60.Gb: Pulsars
% 97.60.Jd: neutron stars
% 97.80.-d  Binary and multiple stars

%-OLD PACS-------------------------------------------
% 04.30.Db: wave generation and sources,
% 97.60.Lf: black holes,
% 04.25.-g: general approximation
%-FROM GOPU------------------------------------------
% from Gopu:
% \pacs{04.30Db,04.25.Nx,04.80.Nn,95.55.Ym}

\maketitle

\section{Introduction}
\label{IntroSec}

A detailed knowledge about the dynamics of spinning compact binaries
is of interest to both astrophysicists and theoretical physicists.
The astrophysical interest arises from the highly accurate
radio observations of relativistic binary pulsars, which
can be used to test general relativity in
strong field regimes \cite{Stairs2003}.
Moreover, inspiralling spinning compact binaries are promising
sources of gravitational radiation
for the ground based and proposed space based
laser interferometric gravitational-wave detectors \cite{IFs}.
The theoretical interest is due to the
possibility of constructing numerically
spinning compact binary configurations
in full general relativity
[see Ref.~\cite{Ansorg05} and references therein].
Further, it is of some interest to explore if the dynamics of
spinning compact binaries will be 
deterministic or not \cite{JL03,GK05nochaos}.

In many of the above mentioned scenarios, the dynamics of 
spinning compact binaries can be described in great detail by the 
post-Newtonian (PN) approximation to general relativity.
The PN approximation allows one to express the equations of motion
of a compact binary as corrections to Newtonian equations of motion
in powers of $(v/c)^2 \sim GM/(c^2 R)$,
where $v$, $M$, and $R$ are
the characteristic orbital velocity,
the total mass and
the typical orbital separation of the binary.
For spinning compact binaries, the dynamics is determined,
not only by the orbital equations of motion for these objects,
but also by the precessional equations for the orbital plane and
the spin vectors themselves \cite{Schaefer_grav_effects_2004}.
Note that for spinning compact binaries, spin effects enter the
dynamics for the first time through spin-orbit interaction.
Therefore, the Hamiltonian 
describing the leading order spin-orbit interaction
can be added to the PN accurate conservative Hamiltonian
which gives the dynamics of nonspinning compact binaries.

It is highly desirable to obtain Keplerian-type parametric solution to
the conservative dynamics of spinning compact binaries.
This is so as 1PN accurate quasi-Keplerian parametrization,
obtained by Damour and Deruelle \cite{DD85}, is heavily employed to
analyze radio observations of relativistic binary pulsars 
and to test general relativity in strong field regimes \cite{DD86,DT92}.
Moreover, 2PN accurate Keplerian-type parametric solution
for nonspinning compact binaries, obtained in Refs.~\cite{DS88,SW93},
is heavily employed to construct highly efficient
``ready to use'' search templates
for nonspinning compact binaries moving in inspiralling
eccentric orbits \cite{DGI}.

Recently, 3PN accurate generalized quasi-Keplerian parametrization
for the conservative dynamics of spinning compact binaries
in eccentric orbits was obtained in Refs.~\cite{MGS,KG05spinorbit}.
More precisely, the PN accurate parametric solutions are
available for the following two distinct cases:
(i) the binary is composed of two compact objects of equal mass and
two arbitrary spins, and
(ii) the two objects have unequal masses but only one object has a spin.
The PN accurate parametric solution may be used to describe 
the following two astrophysical situations.
As the measured neutron star (NS) masses in NS--NS binaries
usually cluster around $1.4 M_{\odot}$,
case (i) may be used to describe such NS--NS binaries.
The case (ii) may be applicable in the instance of a neutron star
orbiting a much larger, rapidly rotating black hole.
In that case, the black-hole spin will
dominate the orbital precession,
and the neutron star spin may be neglected. 

The above parametrization,
for the two cases considered, 
shows that the dynamics is integrable
and cannot be chaotic \cite{GK05nochaos}.
The parametric solution will be employed 
to construct ``ready to use'' search templates 
of spinning compact binaries in inspiralling orbits \cite{KG05_so_phasing}.
Further, it will also  be useful to construct a fully
2PN accurate ``timing formula'', required 
to analyze binary pulsar observations by the
Square Kilometre Array (SKA) \cite{SKAref2004}.

In this paper, we compute, using our parametrization, 
the observable dimensionless relativistic 
periastron advance parameter $k$,
relevant for binary pulsars.
The expression for $k$ that we derive will be applicable, like
the parametrization computed in Ref.~\cite{KG05spinorbit}, only
for the two distinct cases mentioned above.
Our expression for $k$ is in agreement with the one obtained
by Damour and Sch\"afer \cite{DS88}. 
It should be noted that Damour and Sch\"afer employed the
evolution equation for the Laplace-Runge-Lenz vector,
computed from a Hamiltonian, to obtain the spin-orbit contributions
to the parameter $k$. In this paper, we employ our
generalized quasi-Keplerian parametrization,
derived in Ref.~\cite{KG05spinorbit},
to compute the parameter $k$.

There are two reasons that motivated us to do the current computation.
The 2PN accurate expression for $k$, derived by Damour and Sch\"afer, 
valid for arbitrary masses and spins, 
is extremely important due to the discovery
of the double pulsar PSR J0737--3039 \cite{Burgay03}.
This is because the accurate timing of the double pulsar
over a period of few years should lead to the
measurement of the moment of inertia of one of the pulsars \cite{Lyne04}.
It is from the measurement of $k$ to 2PN order, 
which includes leading order spin-orbit contributions,
that one will be able to measure the moment of inertia
of the fast rotating pulsar in this double pulsar \cite{DS88}.
The measurement of the moment of inertia
of a neutron star for the first time will lead to strong constrains
on its equation of state \cite{MBSP04,Lattimer_Schutz_2004}.
We compute $k$, as mentioned earlier,
using Keplerian-type parametrization with the aim of
presenting an alternative way of obtaining this observable quantity.

A second reason for computing $k$ is the existence of certain
intriguing features appearing in our parametrization.
We observed that the factor that modulates the true anomaly, due to
spin-orbit interaction, contains a contribution, independent
of the spins themselves, and of 1PN order
[see Eq.~(5.6k) in Ref.~\cite{KG05spinorbit}].
We want to make sure that this feature will not appear in the 
observable parameter $k$, computed using our
generalized quasi-Keplerian parametrization.

The paper is structured in the following manner. 
In Sec.~\ref{Sec2}, 
we briefly describe the 3PN accurate Keplerian-type parametrization
for the conservative dynamics of spinning compact binaries,
moving in eccentric orbits, 
when spin effects are restricted to the leading order 
spin-orbit interaction, derived in Ref.~\cite{KG05spinorbit},
and point out its salient features.
Sec.~\ref{Sec3} and Sec.~\ref{Sec4} deal with the issues
related to the derivation of the parameter $k$.
We show for the first time how to
derive $k$ using the parametrization, described in Sec.~\ref{Sec2},
after describing how Damour and Sch\"afer obtained $k$.
We summarize our results in Sec.~\ref{SecConclusions}.
Appendices \ref{appendix_binary_geometry} and
\ref{appendix_LRL_vector} provide details of our 
computation and consistency checks.
%Appendix~\ref{appendix_binary_geometry} summarizes some 
%aspects of the binary geometry and the related dynamics, and
%in Appendix~\ref{appendix_LRL_vector} we present an
%alternative way via the Laplace-Runge-Lenz vector to check
%our earlier results.

%%%%%%%%%%%%%%%%%%%%%%%%%%%%%%%%%%%%%%%%%%%%%%%%%%%%%%%%%%%%%%%%%%%%%%%%%%
%%%%%%%%%%%%%%%%%%%%%%%%%%%%%%%%%%%%%%%%%%%%%%%%%%%%%%%%%%%%%%%%%%%%%%%%%%

\section{Generalized quasi-Keplerian parametrization
for spinning compact binaries}
%\section{Third PN Dynamics of compact binaries with leading order
%spin-orbit interaction and its parametrization}  
\label{Sec2}

Recently, 3PN accurate generalized quasi-Keplerian
parametrization for compact binaries in eccentric orbits
with the first-order spin-orbit interaction was 
obtained in Refs.~\cite{MGS,KG05spinorbit} for 
the following two distinct cases:
(i) the binary is composed of two compact objects of equal mass and
two arbitrary spins, and
(ii) the two objects have unequal masses but only one object has a spin.
The underlying dynamics is fully specified by the following
PN accurate (reduced) Hamiltonian $H$, written symbolically as
\begin{align}
\label{eq:H_symbolically}
{H}(\vek{r}, \vek{p}, \vek{S}_{1}, \vek{S}_{2}) &
= {H}_{\rm N}(\vek{r}, \vek{p})
+ {H}_{\rm 1PN}(\vek{r}, \vek{p})
+ {H}_{\rm 2PN}(\vek{r}, \vek{p})
\nonumber
\\
& \quad 
+ {H}_{\rm 3PN}(\vek{r}, \vek{p})
+ {H}_{\rm SO}(\vek{r}, \vek{p}, \vek{S}_{1}, \vek{S}_{2})
\,,
\end{align}
where ${H}_{\rm N}$, ${H}_{\rm 1PN}$, ${H}_{\rm 2PN}$, and 
${H}_{\rm 3PN}$ are, respectively, the Newtonian, first, second and
third PN contributions to the conservative dynamics of
compact binaries, when the spin effects are neglected.
The leading order spin-orbit coupling to the binary dynamics is given
by ${H}_{\rm SO}$. 
In the above equation, $\vek{r} = \vek{\cal R}/(G M)$, $r= |\vek{r}|$, 
and $\vek{p} = \vek{\cal P}/\mu$, where
$\vek{\cal R}$ and $\vek{\cal P}$ are the relative separation vector and
its conjugate momentum vector, respectively.
The familiar symbols $M$ and $\mu$ have the usual meaning, namely 
the total mass and the reduced mass.
The explicit expressions for PN corrections, associated with the motion of 
nonspinning compact binaries, were obtained in 
Refs.~\cite{S85_annphys,DS88}.
The spin-orbit contribution, available in 
Refs.~\cite{KG05spinorbit,DS88,Damour2001}, reads
\begin{align}
{H}_{\rm SO}(\vek{r}, \vek{p}, \vek{S}_{1}, \vek{S}_{2})
&= \frac{1}{c^2 r^3} (\vek{r} {\times} \vek{p}) \cdot \vek{S}_\text{eff}
\,,
\end{align}
where the effective spin $\vek{S}_\text{eff}$ is defined by
\begin{align}
\vek{S}_\text{eff}
\equiv \delta_{1} \vek{S}_{1} + \delta_{2} \vek{S}_{2}
\,,
\end{align} 
with 
\begin{subequations}
\begin{align}
\delta_{1} &
%= 2 \eta \left( 1 + \frac{3 m_2}{4 m_1} \right)
= \frac{\eta}{2} + \frac{3}{4} \left(1 - \sqrt{1 - 4 \eta} \right)
\,,
\\
\delta_{2} &
%= 2 \eta \left( 1 + \frac{3 m_1}{4 m_2} \right)
= \frac{\eta}{2} + \frac{3}{4} \left(1 + \sqrt{1 - 4 \eta} \right)
\,.
\end{align}
\end{subequations}
The finite mass ratio $\eta$ is given by $\eta = \mu /M$.
The (reduced) spin vectors $\vek{S}_{1}$ and $\vek{S}_{2}$ are related
to the individual spins $\vek{\cal S}_1$ and $\vek{\cal S}_2$ by 
$\vek{S}_{1} = \vek{\cal S}_1 /(\mu G M)$ and
$\vek{S}_{2} = \vek{\cal S}_2 /(\mu G M)$, respectively.
We recall that $\vek{\cal R}$, $\vek{\cal P}$, $\vek{\cal S}_1$, and
$\vek{\cal S}_2$ are canonical variables, such that the orbital 
variables commute with the spin variables, e.g.,
see Refs.~\cite{Schaefer_grav_effects_2004,Damour2001}.

\begin{figure*}[ht]
  \centering
  \psfrag{X}{$\vek{e}_X = \vek{p}$}
  \psfrag{Y}{$\vek{e}_Y $}
  \psfrag{J}{$\vek{J}= J \vek{e}_Z$}
  \psfrag{L}{$\vek{L}= L \vek{k}$}
  \psfrag{S}{$\vek{S}$}
  \psfrag{alpha}{$\alpha$}
  \psfrag{N}{$\vek{N}$}
  \psfrag{q}{$\vek{q}$}
  \psfrag{n}{$\vek{n} = \vek{r} / r$}
  \psfrag{a}{$\vek{a}$}
  \psfrag{b}{$\vek{b}$}
  \psfrag{iTheta}{$\vek{i}$}
  \psfrag{ii}{$\vek{i}_{i}$}
  \psfrag{j}{$\vek{j}$}
  \psfrag{i}{$i$}
  \psfrag{i0}{$i_{0}$}
  \psfrag{phi3}{$\Pi$}
  \psfrag{varphi}{$\varphi$}
  \psfrag{v}{$v$}
  %\psfrag{theta1}{$\theta$}
  \psfrag{theta2}{$\Theta$}
  \psfrag{Upsilon}{$\Upsilon$}
  \psfrag{omega}{$\omega$}
  \psfrag{Omega}{$\Omega$}
  \psfrag{Delta}{$\Delta$}
  %\psfrag{lineofsight}{\small{Line of sight}}
  \psfrag{peri}{\small{Periastron}}
  \psfrag{invariable}{\small{Invariable plane}}
  \psfrag{orbit}{\textbf{\small{Orbital plane}}}
  \psfrag{sky}{\small{Plane of the sky}}
  \includegraphics[scale=0.8]{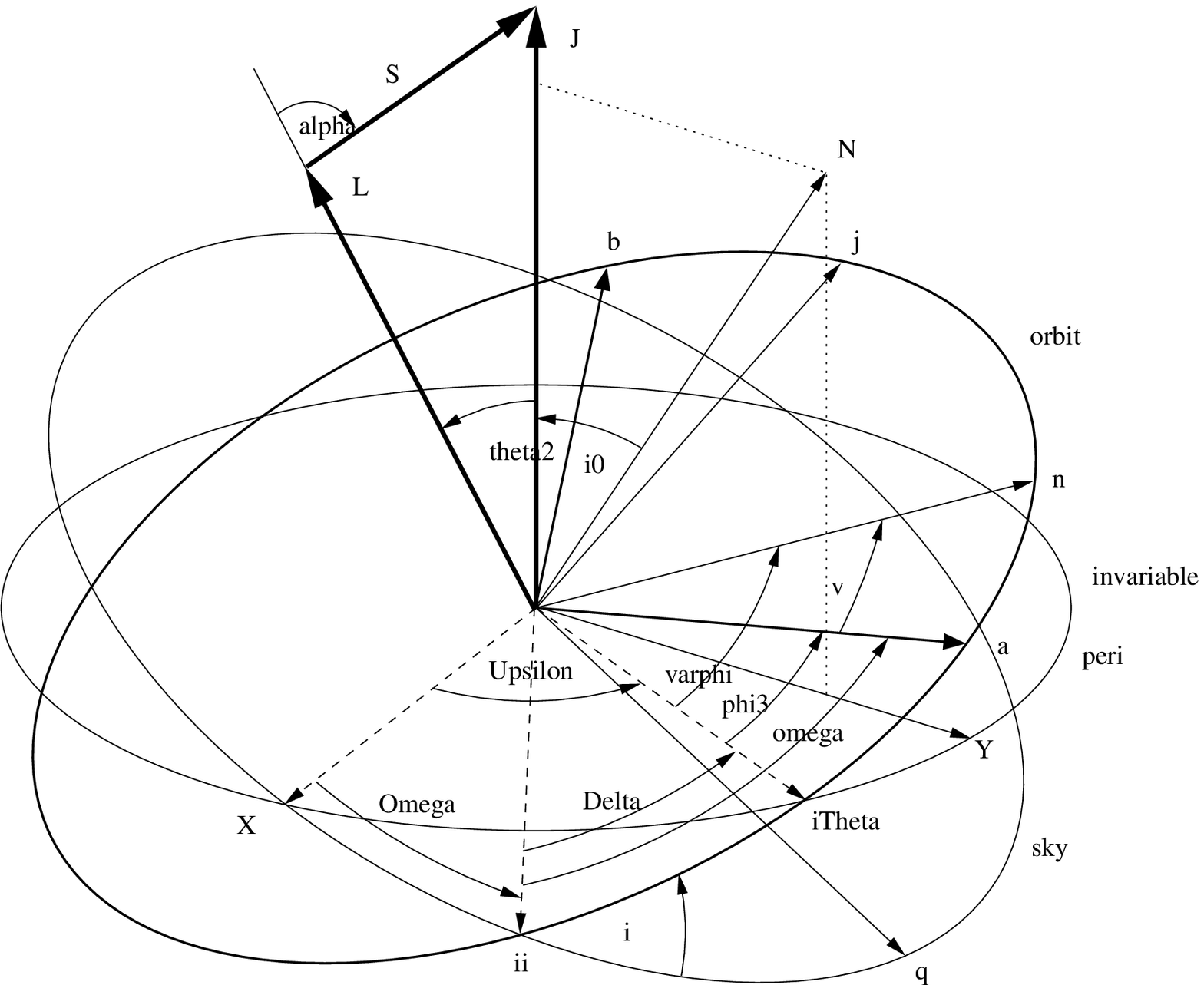}
  \caption{ 
  The binary geometry and definition of various angles.
  Our reference frame is ($\vek{e}_{X}, \vek{e}_{Y},\vek{e}_{Z}$),
  where the basic vector $\vek{e}_{Z}$ is aligned with the fixed total
  angular momentum vector $\vek{J} = \vek{L} + \vek{S}$,
  such that $\vek{J} = J \vek{e}_Z$.
  The invariable plane ($\vek{e}_{X}$, $\vek{e}_{Y}$) is
  perpendicular to $\vek{J}$. Important for the observation is the
  line-of-sight unit vector $\vek{N}$,
  pointing from the source (compact binary) to the observer. 
  We may have, by a clever choice of $\vek{e}_{X}$ and $\vek{e}_{Y}$, the
  line-of-sight unit vector $\vek{N}$ in the
  $\vek{e}_{Y}$-$\vek{e}_{Z}$ plane. 
  $\vek{k} = \vek{L}/L$ is the unit vector in the direction of the
  orbital angular momentum $\vek{L}$, which is perpendicular to the
  orbital plane. The constant inclination of the orbital plane with
  respect to the invariable plane is $\Theta$, which is also the
  precession cone angle of $\vek{L}$ around $\vek{J}$.
  The three lines of nodes are associated with the intersections of the 
  orbital plane, the invariable plane and the plane of the sky.
  The orbital plane intersects the invariable plane at the line of
  nodes $\vek{i}$, with longitude $\Upsilon$, measured in the
  invariable plane from $\vek{e}_{X} = \vek{p}$, which also gives
  the phase of the orbital plane precession.
  The orbital plane intersects the plane of the sky at the line of
  nodes $\vek{i}_{i}$, with longitude $\Omega$, 
  measured in the plane of the sky, from $\vek{e}_{X} = \vek{p}$. 
  The subscript $i$ in $\vek{i}_{i}$ results from the corresponding 
  \emph{orbital} inclination angle $i$.
  The \emph{dynamical} longitude of periastron, 
  measured in the orbital plane from $\vek{i}$, is $\Pi$.
  The \emph{observational} longitude of periastron, measured in the
  orbital plane from $\vek{i}_{i}$, is $\omega$.
  The above figure implies $\omega = \Delta + \Pi $.}
  \label{fig:3fullplanes} 
\end{figure*}

In Ref.~\cite{KG05spinorbit}, we obtained the Keplerian-type
parametrization in an orbital triad $(\vek{i},\vek{j},\vek{k})$.
In this triad, the unit vector $\vek{i}$
is along the line of nodes, associated
with the intersection of the orbital
plane with the invariable plane $(\vek{e}_{X},\vek{e}_{Y})$,
which is the plane perpendicular to the total (reduced) angular
momentum $\vek{J}$, the only constant vector in our analysis.
The unit vector $\vek{k}$ is defined by $\vek{k} \equiv \vek{L}/L$,
where $\vek{L}=\vek{r} \times \vek{p}$ is the (reduced) orbital 
angular momentum vector, and hence $\vek{k}$ is perpendicular to 
the orbital plane (see Fig.~\ref{fig:3fullplanes}). 
We display below, in its entirety, the parametric solution 
to the conservative 3PN accurate
dynamics of spinning compact binaries moving in eccentric orbits, 
in ADM-type coordinates, with spin effects 
restricted to the leading order spin-orbit interactions.
The dynamical vectors of the problem are given by 
\begin{align}
\label{eq:solution_for_r_3PN_plus_SO}
\vek{r}(t) 
&= r(t) \cos\varphi(t) \, \vek{i}(t) 
+ r(t) \sin\varphi(t) \, \vek{j}(t)
\,,
\\
\vek{L}(t) &= L \vek{k}(t)
\,,
\\
\vek{S}(t) &= J \vek{e}_{Z} - L \vek{k}(t)
\,.
\end{align}
We note that in the equal-mass case (i)
the (reduced) total spin $\vek{S}$ is the sum of two arbitrary spins 
$\vek{S}_{1}$ and $\vek{S}_{2}$, and in the single-spin case (ii) 
$\vek{S}$ is either $\vek{S}_{1}$ \emph{or} $\vek{S}_{2}$.
In our restricted cases, namely (i) and (ii),
the (reduced) orbital angular momentum $\vek{L}$ 
and the (reduced) total spin $\vek{S}$
precess about the conserved (reduced) total angular momentum vector 
$\vek{J} = \vek{L} + \vek{S}$ at the same rate \cite{KG05spinorbit}.
This configuration is called \emph{simple precession} in the literature,
e.g., see Ref.~\cite{Cutler1994}.

The time dependent basic vectors $(\vek{i},\vek{j},\vek{k})$ are
explicitly given by 
\begin{subequations}
\label{eq:ijk_in_the_combined_sol}
\begin{align}
\vek{i}(t) & =
\cos\Upsilon(t)  \vek{e}_{X}
+ \sin\Upsilon(t)  \vek{e}_{Y}
\,,
\\
\vek{j}(t) & =
- \cos\Theta \sin\Upsilon(t)  \vek{e}_{X}
+ \cos\Theta \cos\Upsilon(t)  \vek{e}_{Y}
+ \sin\Theta \, \vek{e}_{Z}
\,,
\\
\vek{k}(t) & =
\sin\Theta \sin\Upsilon(t)  \vek{e}_{X}
-\sin\Theta \cos\Upsilon(t) \vek{e}_{Y}
+\cos\Theta \, \vek{e}_{Z}
\,.
\end{align}
\end{subequations}
The angle $\Theta$ denotes the constant precession cone angle of
$\vek{L} = L \vek{k}$ around $\vek{J}$ and is given by
\begin{subequations}
\label{eq:theta_in_terms_of_SLJ}
\begin{align}
\sin\Theta &= \frac{S \sin\alpha}{J}
\,,
\\
\cos\Theta &= \frac{L + S \cos\alpha}{J}
\,,
\end{align}
\end{subequations}
where the magnitude of the total angular momentum is given by
$J = (L^2 + S^2 + 2 L S \cos\alpha)^{1/2}$ and $\alpha$ is the constant
angle between $\vek{L}$ and $\vek{S}$.

The time evolution for $r(t)$, $\varphi(t)$, and $\Upsilon(t)$ is given in
a parametric and PN accurate form, which reads
\begin{widetext}
\begin{subequations}
\label{eq:FinalParamADM_3PN_plus_SO}
\begin{align}
r &= a_r \left( 1 -e_r \cos u \right )
\,,
\\
\label{eq:kepler_eq_in_combined_solution}
l \equiv n \left( t - t_0 \right)
&= u - e_t \sin u + \left( \frac{g_{4t}}{c^4}
+ \frac{g_{6t}}{c^6} \right) (v - u)
+ \left( \frac{f_{4t}}{c^4} + \frac{f_{6t}}{c^6} \right) \sin v
+ \frac{i_{6t}}{c^6} \sin 2 v
+ \frac{h_{6t}}{c^6}  \sin 3 v
\,,
\\
\label{eq:varphi_min_varphi0_in_the_comb_sol}
\varphi - \varphi_{0}
&= \left( 1 + \tilde{k} \right) v
+ \left( \frac{f_{4\varphi}}{c^4}
+ \frac{f_{6\varphi}}{c^6} \right) \sin 2 v
+ \left( \frac{g_{4\varphi}}{c^4}
+ \frac{g_{6\varphi}}{c^6} \right) \sin 3 v
+ \frac{i_{6\varphi}}{c^6} \sin 4 v
+ \frac{h_{6\varphi}}{c^6} \sin 5 v
\,,
\\
\label{eq:Ups_min_Ups0_in_the_comb_sol}
\Upsilon - \Upsilon_0 &= \frac{\chi_\text{so} J}{c^2 L^3} ( v + e \sin v )
\,,
\\
\text{where} \quad
v &= 2\arctan \left[ \left( \frac{ 1 + e_{\varphi}}{ 1 - e_{\varphi}}
\right)^{1/2} \tan \frac{u}{2} \right]
\,.
\end{align}
\end{subequations}
%\end{widetext}
The PN accurate expressions for the orbital elements
$a_r$, $e_r^2$, $n$, $e_t^2$, $\tilde{k}$, and $e_{\varphi}^2$
and the PN orbital functions
$g_{4t}$, $g_{6t}$, $f_{4t}$, $f_{6t}$, $i_{6t}$, $h_{6t}$, $f_{4\varphi}$,
$f_{6\varphi}$, $g_{4\varphi}$, $g_{6\varphi}$, $i_{6\varphi}$,
and $h_{6\varphi}$, in terms of $E$, $L$, $S$, $\eta$, and $\alpha$,
are available in Ref.~\cite{KG05spinorbit}.
The mass dependent coupling constant $\chi_\text{so}$ connects
the effective spin with the total (reduced) spin via 
$\vek{S}_\text{eff}
= \delta_{1} \vek{S}_{1} + \delta_{2} \vek{S}_{2}
= \chi_\text{so} \vek{S}$
and is given by
\begin{equation}
\label{eq:coupling_constant_chi_so}
\chi_\text{so} \equiv
\begin{cases}
\delta_{1} = \delta_{2} = 7/8
& \text{for (i), the equal-mass case}
\,,
\\
\delta_{1}(\eta) \; \text{or} \; \delta_{2}(\eta)
& \text{for (ii), the single-spin case}
\,.
\end{cases}
\nonumber
\end{equation}
We display below the 3PN accurate expression for the 
\emph{internal} parameter $\tilde{k}$,
appearing in Eq.~\eqref{eq:varphi_min_varphi0_in_the_comb_sol}
[this parameter was denoted by $k$ in Ref.~\cite{KG05spinorbit},
see Eqs.~(5.5c) and (5.6k) there].
It consists of two parts:
the first one due to the 3PN accurate conservative
dynamics of nonspinning point masses and
the second one due to the leading order spin-orbit interaction
%\begin{widetext}
\begin{align}
\label{eq:parameter_k_3PN_with_SO}
\tilde{k} = \tilde{k}_\text{3PN} + \tilde{k}_\text{so}
\,,
\end{align}
where
\begin{subequations}
\label{eq:parameter_k_3PN_and_k_so}
\begin{align}
\label{eq:just_parameter_k_3PN}
\tilde{k}_\text{3PN}
&= \frac{3}{c^2 L^2}
\biggl\{
1 
+ \frac{ (-2 E) }{4 c^2}
\left( - 5 + 2 \eta + \frac{ 35 - 10 \eta }{ -2 E L^2 } \right)
+ \frac{ (-2 E)^2 }{ 384 c^4 } 
\biggl[
120 - 120 \eta + 96 \eta^2 
\nonumber
\\
& \quad
+ \frac{1}{ ( - 2 E L^2) }
( - 10080 + 13952 \eta - 123 \pi^2 \eta - 1440 \eta^2 )
\nonumber
\\
& \quad
+ \frac{1}{ (- 2 E L^2)^2 } 
( 36960 - 40000 \eta +  615 \pi^2 \eta + 1680 \eta^2 )
\biggr]
\biggr\}
\,,
\\
\label{eq:just_parameter_k_so}
\tilde{k}_\text{so}
&= \frac{3}{c^2 L^2}
\biggl\{
- \frac{ \chi_\text{so} }{3} - \chi_\text{so} \frac{S}{L} \cos\alpha
\biggr\}
\,.
\end{align}
\end{subequations}
\end{widetext}
It should be noted that $\tilde{k}_\text{3PN}$ gives
the observable advance of periastron in the case of nonspinning
compact binaries. This results from the construction of $\tilde{k}$
[see Sec.~V in Ref.~\cite{KG05spinorbit} for details 
and Ref.~\cite{DS88}].
However, when spin-orbit interactions are added to the dynamics of
compact binaries, $\tilde{k}$ does not directly lead 
to the measurable advance of periastron,
relevant for binary pulsars.
This parameter $\tilde{k}$ is a measure for the advance of periastron
\emph{only inside} our parametrization and has to be related to
the observational one.
Further, note the presence of the $\chi_\text{so}$ term
(which does not depend on the spins),
appearing in $\tilde{k}_\text{so}$ at the 1PN order.
It is due to these points, namely 
the $\chi_\text{so}$ term that appears at the 1PN order and 
$\tilde{k}$ does not give the observable periastron
advance for spinning compact binaries, that
we compute the dimensionless observable
periastron advance parameter $k$ for spinning compact binaries
in this paper.
Note, that we have corrected the sign in front 
of $\chi_\text{so} / 3$ in $\tilde{k}_\text{so}$, important for
the proper nonspinning limit and the following calculation of the
observable periastron advance parameter $k$ in Sec.~\ref{Sec4}.

Before we go onto the details of computing $k$, let us list a few
salient featurs of the parametrization.
The parameter $e$, appearing in
Eq.~\eqref{eq:Ups_min_Ups0_in_the_comb_sol}, is given by the 
Newtonian contribution to say $e_r$. 
Note that at Newtonian order all three eccentricities
$e_r$, $e_t$, and $e_\varphi$, are identical to each other and
at PN orders they are related by PN accurate relations
[see Eqs.~(5.7) in Ref.~\cite{KG05spinorbit}].
Therefore, it is possible to describe the binary dynamics in terms 
of one of the eccentricities.
We emphasize that, while adapting the parametrization to describe the
binary dynamics, care should be taken to restrict orbital elements to
the required PN order. 
We are aware that the above parametrization does not lead 
to the classic Keplerian parametrization simply by putting $S = 0$.
The detailed explanation of the nonspinning limit and the 
apparent lack of a \emph{simple} nonspinning limit is given in 
Sec.~IV~B in Ref.~\cite{KG05spinorbit}.

The above parametric solution is usually refered to as the
genaralized quasi-Keplerian parametrization for the PN accurate 
orbital dynamics of spinning compact binaries.
It extends the computations of Damour, Deruelle, Sch\"afer, and Wex
\cite{DD85,DS88,SW93,MGS,Wex1995}.

Finally, we want to emphasize that the parametric solution,
given by Eqs.~\eqref{eq:solution_for_r_3PN_plus_SO}--\eqref{eq:parameter_k_3PN_and_k_so},
describes the entire conservative PN accurate
dynamics of a spinning compact binary in an eccentric orbit, 
when the spin effects are restricted to the leading order 
spin-orbit interaction. This means the parametrization consistently
describes not only the precessional motion of the orbit
inside the orbital plane, but also the precessional motions 
of the orbital plane and the spins themselves.
Further, the deterministic nature of the 
dynamics described by the Hamiltonian,
given by Eq.~\eqref{eq:H_symbolically} and treated as a
self-contained dynamical system, naturally follows
from the above parametric solution.

In the next section, we develop the basics required to compute
$k$ from our parametrization.

%%%%%%%%%%%%%%%%%%%%%%%%%%%%%%%%%%%%%%%%%%%%%%%%%%%%%%%%%%%%%%%%%%%%%%%%%%
%%%%%%%%%%%%%%%%%%%%%%%%%%%%%%%%%%%%%%%%%%%%%%%%%%%%%%%%%%%%%%%%%%%%%%%%%%

\section{Relevant inputs required to compute $k$}
%\section{Contact with the observation}
\label{Sec3}

In this section, we will introduce two new triads, in terms 
of which it will be easy to obtain $k$. We will also
discuss in detail how to connect these triads and hence
the parametrization expressed in terms of these triads.

%%%%%%%%%%%%%%%%%%%%%%%%%%%%%%%%%%%%%%%%%%%%%%%%%%%%%%%%%%%%%%%%%%%%%%%%%%

\subsection{The periastron-based triad for the parametrization}
%\label{}
As we are interested in computing $k$ using our parametrization,
it is natural that we introduce a periastron-based triad
$(\vek{a},\vek{b},\vek{k})$.
Later, we will express the relative separation vector $\vek{r}$,
given by Eq.~\eqref{eq:solution_for_r_3PN_plus_SO},
in terms of this new triad.

Let us introduce this periastron-based triad in a pedagogical way.
We start by writing the angle $\varphi$,
given by Eq.~\eqref{eq:varphi_min_varphi0_in_the_comb_sol},
in the following way
\begin{align}
\label{eq:ansatz_for_varphi}
\varphi &= v + \Pi
\,.
\end{align}
Remember, $\varphi$ is the polar angle inside
the orbital plane measured from the line of nodes $\vek{i}$, 
associated with the intersection of the orbital plane
with the invariable plane $(\vek{e}_{X},\vek{e}_{Y})$, 
and $v$ is the ``true anomaly'', i.e., the angle between the
(instantaneous) periastron and the position of 
the relative separation vector $\vek{r}$.
The angle $\Pi$ may be considered as the
``argument of the periastron'', i.e.,
the angle between the line of nodes $\vek{i}$
and the (instantaneous) periastron.
By comparing Eqs.~\eqref{eq:ansatz_for_varphi} and
\eqref{eq:varphi_min_varphi0_in_the_comb_sol},
we can easily infer the parametric evolution for $\Pi$.
Therefore, it is obvious that the angle $\Pi$ contains 
via $\tilde{k}$ the secular contributions to the periastron advance.

Inserting Eq.~\eqref{eq:ansatz_for_varphi} into 
Eq.~\eqref{eq:solution_for_r_3PN_plus_SO}
and making use of familiar trigonometric
relations leads to
\begin{align}
\label{eq:r_in_terms_of_Pi_and_v}
\vek{r} 
&= r \cos(v + \Pi) \, \vek{i} + r \sin(v + \Pi) \, \vek{j}
\nonumber \\
&= r \cos v \, \vek{a} + r \sin v \, \vek{b}
\,,
\end{align}
where the new basic vectors are given by
\begin{subequations}
\begin{align}
\vek{a} &= \cos \Pi \, \vek{i} + \sin \Pi \, \vek{j} \,, \\
\vek{b} &= - \sin \Pi \, \vek{i} + \cos \Pi \, \vek{j} \,.
\end{align}
\end{subequations}
Now, it is obvious that $\vek{a}$ is the unit vector pointing
towards the (instantaneous) periastron, as $\vek{r}$ is periodic
in the angular variable $v$ with period $2 \pi$ in these new 
basic vectors $\vek{a}$ and $\vek{b}$. 
Hence it is straightforward to define the periastron-based orthonormal triad
$(\vek{a},\vek{b},\vek{k}) \equiv (\vek{a},\vek{k} \times \vek{a},\vek{k})$, 
by rotating the $(\vek{i},\vek{j},\vek{k})$ frame around the orbital
angular momentum unit vector $\vek{k}$, by an angle $\Pi$.
This gives us
\begin{align}
\label{eq:def_abk_from_ijk}
\begin{pmatrix}
\vek{a}\\
\vek{b}\\
\vek{k}
\end{pmatrix}
&=
\begin{pmatrix}
\cos\Pi &\sin\Pi &0 \\
-\sin\Pi &\cos\Pi &0 \\
0 &0 &1
\end{pmatrix}
%-----------------------
\begin{pmatrix}
\vek{i} \\
\vek{j} \\
\vek{k}
\end{pmatrix}
\,,
\end{align}
where $\Pi$ is the angle between $\vek{i}$ and the
(instantaneous) periastron unit vector $\vek{a}$, as explained above.

The periastron-based vectors $(\vek{a},\vek{b})$ 
are co-rotating with the precessing (instantaneous)
Newtonian ellipse. This is similar to the behaviour of a body fixed
reference system 
usually encountered in classical mechanics \cite{Goldstein}.
Hence, the set of angles $(\Upsilon,\Pi,\Theta)$ is
equivalent to the Eulerian angles defined there.

In the next subsection, we will introduce an observer-based
triad and investigate how to connect the observer
and periastron-based triads.

%%%%%%%%%%%%%%%%%%%%%%%%%%%%%%%%%%%%%%%%%%%%%%%%%%%%%%%%%%%%%%%%%%%%%%%%%%

\subsection{Connecting the observer and periastron-based triads}
%\label{}

The second new triad that we introduce is a fixed one associated
with the observer on earth and denoted by $(\vek{p},\vek{q},\vek{N})$.
In this fixed triad $\vek{p}$ and $\vek{q}$ are two orthogonal 
unit vectors in the plane of the sky, i.e., the plane transverse 
to the radial direction $\vek{N}$,
pointing from the source (compact binary) to the observer
(see Fig.~\ref{fig:3fullplanes}). 

In order to compute $k$ using our parametrization, it is natural
to expect that we need to connect 
the periastron-based triad $(\vek{a},\vek{b},\vek{k})$
to the observer triad $(\vek{p},\vek{q},\vek{N})$.
It is possible to connect these two frames
with the help of rotational matrices quite similar to
the way in which the $(\vek{i},\vek{j},\vek{k})$ frame was connected to
the $(\vek{p},\vek{q},\vek{N})$ frame in Ref.~\cite{KG05spinorbit}
[see Sec.~VI in Ref.~\cite{KG05spinorbit} for details].
In this paper, we come from the observer triad $(\vek{p},\vek{q},\vek{N})$
to the periastron-based triad $(\vek{a},\vek{b},\vek{k})$
with the help of \emph{four} successive rotations:
first, a rotation around $\vek{p}$ via the \emph{constant} angle $i_{0}$
to come to the reference triad $(\vek{e}_{X},\vek{e}_{Y},\vek{e}_{Z})$,
second, a rotation around $\vek{e}_Z$ by an angle $\Upsilon$,
third, a rotation around the direction of the line of nodes 
$\vek{i}$ [associated with the intersection of the
orbital plane with the invariable plane $(\vek{e}_{X},\vek{e}_{Y})$]
by an angle $\Theta$, and,
finally fourth, a rotation around the orbital angular momentum direction
$\vek{k}$ by an angle $\Pi$.
In terms of rotational matrices --- denoted by
${\cal R}[\text{rotation angle}, \text{rotation axis}]$ --- we have 
\begin{align}
\label{eq:def_abk_from_exeyez_via_KG2005}
\begin{pmatrix}
\vek{a}(t)\\
\vek{b}(t)\\
\vek{k}(t)
\end{pmatrix}
&= 
{\cal R}[\Pi(t), \vek{k}(t)]       \; 
{\cal R}[\Theta, \vek{i}(t)]       \; 
{\cal R}[\Upsilon(t), \vek{e}_{Z}]  
\nonumber
\\
& \quad \times {\cal R}[i_{0}, \vek{p}]
%-----------------------
\begin{pmatrix}
\vek{p} \\
\vek{q} \\
\vek{N}
\end{pmatrix}
\,,
\end{align}
where the first rotation appears at the right-hand side, 
the second one appears left from the one before, and so on.

We note, that the approach to generate the 
$(\vek{a},\vek{b},\vek{k})$ frame via these
rotations may be also adapted for scenarios, where 
the angle $\Theta$ depends on time. 
The constant behaviour of $\Theta$
is a direct consequence of choosing two distinct cases (i) and (ii),
where our parametrization is valid and hence we will use
the fact that $\Theta$ is constant only at the end of our analysis.

It should be noted that there are other
possible ways to connect the orthonormal triads 
$(\vek{p},\vek{q},\vek{N})$ and $(\vek{a},\vek{b},\vek{k})$.
A second way to generate the $(\vek{a},\vek{b},\vek{k})$ frame 
is described in great detail by Damour and Sch\"afer
in Sec.~5 of Ref.~\cite{DS88} and involves
\emph{three} rotational matrices.
This way is closely linked to the observation
as it involves rotational matrices
defined with the help of the 
observational longitude of periastron $\omega$
and the orbital inclination angle $i$.
Damour and Sch\"afer connected 
the observer triad $(\vek{p},\vek{q},\vek{N})$
to the periastron-based triad $(\vek{a},\vek{b},\vek{k})$
via the following \emph{three} successive rotations:
first, a rotation around $\vek{N}$ by an angle $\Omega$,
second, a rotation around the direction of the line of nodes 
$\vek{i}_{i}$ [associated with the intersection of the
orbital plane with the plane of the sky $(\vek{p},\vek{q})$]
by an angle $i$, and,
finally third, a rotation around the orbital angular momentum direction
$\vek{k}$ by an angle $\omega$. The corresponding
rotational matrices read 
\begin{align}
\label{eq:def_abk_from_pqN_via_DS88}
\begin{pmatrix}
\vek{a}(t)\\
\vek{b}(t)\\
\vek{k}(t)
\end{pmatrix}
&= 
{\cal R}[\omega(t), \vek{k}(t)]  \; 
{\cal R}[i(t), \vek{i}_{i}(t)]   \;
{\cal R}[\Omega(t), \vek{N}]
%-----------------------
\begin{pmatrix}
\vek{p} \\
\vek{q} \\
\vek{N}
\end{pmatrix}
\,.
\end{align}

We note, that for the unit vector $\vek{i}_{i}$ 
along the line of nodes, associated with the intersection of the
orbital plane with the plane of the sky, 
the subscript $i$ is employed as $\vek{i}_{i}$ 
is related to the orbital inclination angle $i$ 
and to avoid any confusion with the unit vector $\vek{i}$. 

In the next section, we obtain $k$ with the 
help of the parametrization
and the ideas discussed in this section.

%%%%%%%%%%%%%%%%%%%%%%%%%%%%%%%%%%%%%%%%%%%%%%%%%%%%%%%%%%%%%%%%%%%%%%%%%%
%%%%%%%%%%%%%%%%%%%%%%%%%%%%%%%%%%%%%%%%%%%%%%%%%%%%%%%%%%%%%%%%%%%%%%%%%%

\section{The advance of periastron relevant for binary pulsars}
\label{Sec4}

In this section, we will compute $k$,
the observable dimensionless periastron advance parameter 
due to $\tilde{k}$.
Following Damour and Sch\"afer, the PN accurate total
observable dimensionless periastron advance parameter $k$
can be written as
\begin{align}
\label{eq:parameter_k_total}
k = k_\text{3PN} + k_\text{so}
\,,
\end{align}
where $k_\text{3PN}$ is the usual PN accurate periastron advance parameter 
up to 3PN order, corresponding to the nonspinning case
[see Refs.~\cite{DS88,MGS} for the 
2PN and 3PN contributions, respectively],
and $k_\text{so}$ denotes the contribution due to spin-orbit interaction.
It should be kept in mind that the 3PN contribution to $k_\text{3PN}$
will not be required while probing the dynamics
of relativistic binary pulsars even with SKA.

Let us first briefly summarize how Damour and Sch\"afer obtained
$k$, especially $k_\text{so}$, in the next subsection.

%%%%%%%%%%%%%%%%%%%%%%%%%%%%%%%%%%%%%%%%%%%%%%%%%%%%%%%%%%%%%%%%%%%%%%%%%%

\subsection{The derivation of $k_\text{so}$ --- The approach 
of Damour and Sch\"afer}

Timing of binary pulsars allows via the observed
``Roemer time delay'' the direct measurement of the 
argument of periastron $\omega$ \cite{DD86}.
Therefore, Damour and Sch\"afer defined 
the relevant \emph{secular} periastron advance as 
$\langle d\omega/dt \rangle_t$, or equivalently the timing parameter
(not to be confused with the orbital angular-momentum unit
vector $\vek{k}$)
\begin{align}
\label{eq:definition_k_obs}
k &\equiv \frac{1}{n} \left\langle \frac{d \omega}{dt} \right\rangle_t
\,,
\end{align}
where $n$ is the mean orbital motion
that appears in the PN accurate Kepler equation,
given by Eq.~\eqref{eq:kepler_eq_in_combined_solution},
and $\langle \ldots \rangle_t$ denotes the orbital average.

Damour and Sch\"afer computed the contribution to $d\omega/dt$ 
due to spin-orbit interaction in the following way.
First, they computed the precessional equations
for the orbital angular momentum vector $\vek{\cal L}$
and the Laplace-Runge-Lenz vector $\vek{\cal A}$,
which tracks the periastron advance,
with the help of the 2PN accurate version of 
Eq.~\eqref{eq:H_symbolically}.
These precessional equations where orbital averaged to obtain
the rotational velocity vector, denoted by $\vek{\Omega}$
in Ref.~\cite{DS88}.
The expression for $d\omega/dt$ was extracted from $\vek{\Omega}$
by decomposing $\vek{\Omega}$ along 
three \emph{non-orthogonal} directions 
[see Eq.~(5.11) in Ref.~\cite{DS88}].
In this way a parametric solution to the dynamics of 
spinning compact binaries 
was not required to compute the 2PN accurate measurable
dimensionless periastron advance parameter $k$ in Ref.~\cite{DS88}, 
and hence the formula of Damour and Sch\"afer for $k$
is more general.
 
In the next subsection, we will compute $d\omega/dt$
using our Keplerian-type parametric solution
and the two different ways of connecting 
the $(\vek{p},\vek{q},\vek{N})$ frame to
the $(\vek{a},\vek{b},\vek{k})$ frame.

%%%%%%%%%%%%%%%%%%%%%%%%%%%%%%%%%%%%%%%%%%%%%%%%%%%%%%%%%%%%%%%%%%%%%%%%%%

\subsection{The derivation of $k_\text{so}$ --- The parametric approach}

This subsection deals with the computation of $d\omega/dt$, arising
due to spin-orbit interaction, using our Keplerian-type parametrization.

Following Damour and Sch\"afer, we introduce the 
rotation velocity vector $\vek{\Omega}$,
containing all involved angular velocities,
and extract $d\omega/dt$ from it in the following way.
We have already noted that there are two different ways to connect
the observer-based triad to the pariastron-based one
[see Eqs.~\eqref{eq:def_abk_from_exeyez_via_KG2005} 
and \eqref{eq:def_abk_from_pqN_via_DS88} 
and related discussions there].
The first way,
given by Eq.~\eqref{eq:def_abk_from_exeyez_via_KG2005},
involves the angular variables 
$\Pi(t)$, $\Theta$, $\Upsilon(t)$, and $i_{0}$,
appearing in our parametrization, and
the second way,
given by Eq.~\eqref{eq:def_abk_from_pqN_via_DS88},
involves the angles $\omega(t)$, $i(t)$, and $\Omega(t)$,
where two of them are relevant for the observations,
namely $\omega(t)$ and $i(t)$.
Therefore, we can decompose the 
rotation velocity vector $\vek{\Omega}$ in the following two ways:
\begin{subequations}
\label{eq:vec_Omega_in_both_ways}
\begin{align}
\label{eq:vec_Omega_1st_way}
\vek{\Omega} &
= \frac{d \Pi}{dt} \vek{k}
+ \frac{d \Theta}{dt} \vek{i}
+ \frac{d \Upsilon}{dt} \vek{e}_{Z}
+ \frac{d i_{0}}{dt} \vek{p}
\,,
\\
\label{eq:vec_Omega_2nd_way}
\vek{\Omega} &
= \frac{d \omega}{dt} \vek{k}
+ \frac{d i}{dt} \vek{i}_{i}
+ \frac{d \Omega}{dt} \vek{N}
\,.
\end{align}
\end{subequations}
Note, that the above two different decompositions
are along four/three \emph{non-orthogonal} directions, respectively.
The elementary rotation velocities, 
appearing in Eq.~\eqref{eq:vec_Omega_1st_way},
can be defined with the help of our parametrization,
whereas in Eq.~\eqref{eq:vec_Omega_2nd_way}
we have $d\omega/dt$: the rate of change
of the argument of the periastron.
Further note, that Eq.~\eqref{eq:vec_Omega_2nd_way}
is identical to Eq.~(5.11) of Ref.~\cite{DS88}.

In order to extract $d\omega/dt$ and 
to relate it to the angular variables 
associated with our parametrization we proceed as follows.
Since all vectors, appearing in Eqs.~\eqref{eq:vec_Omega_in_both_ways},
can be expressed in terms of $(\vek{a},\vek{b},\vek{k})$,
it is clear that the rotation velocity vector $\vek{\Omega}$
may also be written as the sum of three elementary rotation
velocities, along three \emph{orthogonal} directions
$\vek{a}$, $\vek{b}$, and $\vek{k}$, namely
\begin{align}
\label{eq:decomposition_Omega}
\vek{\Omega} &= 
\Omega_{(\vek{a})} \vek{a} +
\Omega_{(\vek{b})} \vek{b} +
\Omega_{(\vek{k})} \vek{k}
\,.
\end{align}
It is then straightforward to compute the components
$\Omega_{(\vek{a})}$, $\Omega_{(\vek{b})}$, and $\Omega_{(\vek{k})}$
in terms of the elementary rotation velocities and angles 
appearing in Eqs.~\eqref{eq:vec_Omega_in_both_ways}.
The components of $\vek{\Omega}$ in the co-rotating
periastron-based $(\vek{a},\vek{b},\vek{k})$ frame
--- using the parametrization-based variables
and the observation-based variables, respectively --- 
are given by 
\begin{widetext}
\begin{subequations}
\label{eq:components_of_Omega_in_abk}
\begin{align}
\Omega_{(\vek{a})} &
=\frac{d\Theta}{dt} \cos\Pi
+\frac{d\Upsilon}{dt} \sin\Theta \sin\Pi
+\frac{d i_{0}}{dt} (\cos\Pi \cos\Upsilon - \cos\Theta \sin\Pi \sin\Upsilon) 
=\frac{d i}{dt} \cos\omega
+\frac{d\Omega}{dt} \sin i \sin\omega 
\,, \\
\Omega_{(\vek{b})} &
=-\frac{d\Theta}{dt} \sin\Pi
+\frac{d\Upsilon}{dt} \sin\Theta \cos\Pi 
+\frac{d i_{0}}{dt} (-\sin\Pi \cos\Upsilon - \cos\Theta \cos\Pi \sin\Upsilon) 
=-\frac{d i}{dt} \sin\omega
+\frac{d\Omega}{dt} \sin i \cos\omega 
\,, \\
\Omega_{(\vek{k})} &
=\frac{d \Pi}{dt}
+\frac{d \Upsilon}{dt} \cos\Theta 
+\frac{d i_{0}}{dt} \sin\Theta \sin\Upsilon
=\frac{d \omega}{dt} 
+\frac{d \Omega}{dt} \cos i 
\,.
\end{align}
\end{subequations}
\end{widetext}
These \emph{general} set of equations provides the 
transformation from the parametrized angles
to the angles of interest for the observation.
We note, that the last component, $\Omega_{(\vek{k})}$, gives the 
so-called \emph{intrinsic} contribution to the periastron advance
$\dot{\omega}^\text{intrinsic}$, defined by 
$\dot{\omega}^\text{intrinsic} \equiv \vek{\Omega} \cdot \vek{k}$.
This is the component of $\vek{\Omega}$ acting \emph{within}
the orbital plane. However, in the timing of binary
pulsars the parameter $k$ enters in the (relativistic)
argument of the periastron $\omega$, and so we 
have to extract $d\omega/dt$ from the above set of equations.

A close inspection of Fig.~\ref{fig:3fullplanes} indicates
that $i_{0}$ is constant, since it connects two
invariant reference frames. 
For the two distinct cases (i) and (ii), 
where our parametrization is valid,
we also have $\Theta$ as a constant angle.
This allows us to come from Eqs.~\eqref{eq:components_of_Omega_in_abk}
to the \emph{reduced} set of equations for the components 
$\Omega_{(\vek{a})}$, $\Omega_{(\vek{b})}$, and $\Omega_{(\vek{k})}$,
which reads
\begin{subequations}
\label{eq:reduced_components_of_Omega_in_abk}
\begin{align}
\label{eq:reduced_I}
\frac{d\Upsilon}{dt} \sin\Theta \sin\Pi
&=\frac{d i}{dt} \cos\omega
+\frac{d\Omega}{dt} \sin i \sin\omega 
\,, \\
\label{eq:reduced_II}
\frac{d\Upsilon}{dt} \sin\Theta \cos\Pi 
&=-\frac{d i}{dt} \sin\omega
+\frac{d\Omega}{dt} \sin i \cos\omega 
\,, \\
\label{eq:reduced_III}
\frac{d \Pi}{dt}
+\frac{d \Upsilon}{dt} \cos\Theta 
&=\frac{d \omega}{dt} 
+\frac{d \Omega}{dt} \cos i 
\,.
\end{align}
\end{subequations}
Solving the last equation, namely Eq.~\eqref{eq:reduced_III},
for $d\omega/dt$ leads to
\begin{align}
\label{eq:domega_dt}
\frac{d \omega}{dt} &
= \frac{d \Pi}{dt} + \frac{d\Upsilon}{dt} \cos\Theta 
- \frac{d\Omega}{dt} \cos i 
\,.
\end{align}
The first and the second contribution to $d\omega/dt$
are directly provided by our parametrization. 
We also figured out that they can be obtained 
via the time evolution of the Laplace-Runge-Lenz vector
without any parametrization 
(see Appendix~\ref{appendix_LRL_vector} for details).
The third contribution to $d\omega/dt$ in Eq.~\eqref{eq:domega_dt}, 
namely $(d\Omega/dt)\cos i$, needs a closer inspection.
To obtain $d\Omega/dt$,
we multiply Eq.~\eqref{eq:reduced_I} with $\sin\omega$ and add 
it to the product of Eq.~\eqref{eq:reduced_II} times $\cos\omega$,
which results in
\begin{align}
\frac{d \Omega}{dt} &
= \frac{1}{\sin i} \frac{d \Upsilon}{dt} \sin\Theta 
\left( \sin\Pi \sin\omega + \cos\Pi \cos\omega \right)
\,.
\end{align}
Using now the explicit expressions
for $\sin\omega$ and $\cos\omega$, 
given by Eqs.~\eqref{eq:cos_omega_sin_omega}
in Appendix~\ref{appendix_binary_geometry},
in the above equation, we get 
\begin{align}
\label{eq:dOmega_dt}
\frac{d \Omega}{dt} &
= \frac{1}{\sin^2 i} \frac{d \Upsilon}{dt}
S_\Theta (C_{i_{0}} S_\Theta  + S_{i_{0}} C_\Theta \cos\Upsilon )
\,,
\end{align}
where 
$S_\Theta$, $C_\Theta$, $S_{i_{0}}$, and $C_{i_{0}}$
are shorthand notations for the constant trigonometric functions
$\sin\Theta$, $\cos\Theta$, $\sin i_{0}$, and $\cos i_{0}$,
respectively.
The time evolution for the orbital inclination $i$,
appearing in Eqs.~\eqref{eq:domega_dt}--\eqref{eq:dOmega_dt}, 
is simply given by Eqs.~\eqref{eq:time_evolution_of_i}
in Appendix~\ref{appendix_binary_geometry}.

Let us now explicitly compute $d\Pi/dt$, $d\Upsilon/dt$ 
and hence $d\Omega/dt$ with the help of our parametric solution.
Using Eqs.~\eqref{eq:varphi_min_varphi0_in_the_comb_sol},
\eqref{eq:parameter_k_3PN_with_SO}, and
\eqref{eq:ansatz_for_varphi} for $\Pi$, and
Eq.~\eqref{eq:Ups_min_Ups0_in_the_comb_sol} for $\Upsilon$,
we display below the relevant parametric expressions
symbolically
\begin{subequations}
\label{eq:display_the_separation}
\begin{align}
\Pi - \Pi_0 &
= \left( \tilde{k}_\text{3PN} + \tilde{k}_\text{so} \right) v
+ (\ldots) \sin 2 v
+ (\ldots) \sin 3 v
\nonumber \\
& \quad
+ (\ldots) \sin 4 v
+ (\ldots) \sin 5 v
\,,
\\
\Upsilon - \Upsilon_0 &
= \frac{\chi_\text{so} J}{c^2 L^3} ( v + e \sin v )
\,.
\end{align}
\end{subequations}
The structure of Eqs.~\eqref{eq:display_the_separation} 
indicates that we can handle 
the PN and the spin-orbit induced contributions 
to $d\omega/dt$ separately. 

To obtain the spin-orbit contributions to
$\langle d\omega/dt \rangle_{t}$,
we compute with the help of 
Eqs.~\eqref{eq:display_the_separation} and \eqref{eq:just_parameter_k_so} 
the orbital averaged spin-orbit (so) contributions 
to the above required time derivatives
%\begin{widetext}
\begin{subequations}
\label{eq:so_contr_to_Pi_Upsilon_Omega_REDUCED}
\begin{align}
\label{eq:so_contr_to_the_aver_Pi}
\left \langle 
\frac{ d \Pi }{ d t } 
\right \rangle_{t}^{\rm so} &
= \frac{ 1 }{ c^2 } 
\left \langle\frac{1}{ {r}^3 } \right \rangle_{t}
\chi_{\rm so} 
( - L - 3 S \cos\alpha )
\,,
\\
\left \langle 
\frac{ d \Upsilon }{ d t } C_\Theta
\right \rangle_{t}^{\rm so} &
= \frac{ 1 }{ c^2 } 
\left \langle\frac{1}{ {r}^3 } \right \rangle_{t}
\chi_{\rm so} 
( L + S \cos\alpha )
\,,
\\
\left \langle 
\frac{ d \Omega }{ d t } \cos i
\right \rangle_{t}^{\rm so} &
= \frac{ 1 }{ c^2 } 
\left \langle\frac{1}{ {r}^3 } \right \rangle_{t}
\chi_{\rm so} {S}
\sin\alpha \frac{ \cos i }{ \sin^2 i }
\nonumber
\\
& \quad
\times
(C_{i_{0}} S_\Theta  + S_{i_{0}} C_\Theta \cos\Upsilon )
\,.
\end{align}
\end{subequations}
%\end{widetext}
The quantity $\langle 1/r^3 \rangle_{t}$ gives the orbital average
of $1/r^3$, $r$ being the scaled radial separation, 
to the Newtonian order [see, e.g., Ref.~\cite{DS88}]
\begin{align}
\label{eq:OneOverrcube_averaged}
\left \langle\frac{ 1 }{ {r}^3 } \right \rangle_{t}
& 
= \frac{ 1 }{ a^3_r (1 - e^2)^{3/2} } = \frac{ n^2 }{ (1 - e^2)^{3/2} }
\,.
\end{align}

With the help of 
Eqs.~\eqref{eq:domega_dt},
\eqref{eq:so_contr_to_Pi_Upsilon_Omega_REDUCED}, and
\eqref{eq:OneOverrcube_averaged},
it is straightforward to obtain
the spin-orbit induced contribution to $k$, defined by 
Eqs.~\eqref{eq:parameter_k_total} and \eqref{eq:definition_k_obs},
and it reads
\begin{align}
\label{eq:k_obs_CK}
k_\text{so}
& = B \left[
- 2 \cos\alpha - \sin\alpha \frac{ \cos i }{ \sin^2 i }
(C_{i_{0}} S_\Theta  + S_{i_{0}} C_\Theta \cos\Upsilon )
\right]
\,,
\end{align}
where $B$ is given in terms of our \emph{reduced} variables by
\begin{align}
\label{eq:coefficient_B_CK}
B &
= \frac{ n \chi_\text{so} S }{ c^2 (1 - e^2)^{3/2} } 
\,.
\end{align}
The proper nonspinning limit of
Eqs.~\eqref{eq:k_obs_CK} and \eqref{eq:coefficient_B_CK},
can be easily deduced as when $S \rightarrow 0$, 
$\vek{J}$ is identical to $\vek{L}$, which implies 
$\Theta \rightarrow 0$ and $i_{0} \rightarrow i$, 
$i$ being the angle between $\vek{N}$ and $\vek{L}$,
and therefore $k_\text{so}$ vanishes.
The disappearance of $k_\text{so}$ as $S \rightarrow 0$ indicates
that the presence of the 1PN accurate
$\chi_\text{so}$ term in $\tilde{k}_\text{so}$, 
given by Eq.~\eqref{eq:just_parameter_k_so},
does not have any observational consequence.

Our result for $k_\text{so}$ is in agreement with 
Eqs.~(5.17) and (5.17a) of Ref.~\cite{DS88},
when we specialize these equations
to the \emph{simple precession}
(for the equal-mass case and for the single-spin case)
\begin{align}
\label{eq:k_obs_DS88}
k_\text{so}
& = \frac{ B }{ \sin^2 i } 
\left[ ( 1 - 3 \sin^2 i ) (\vek{k} \cdot \vek{s} ) 
- \cos i (\vek{N} \cdot \vek{s} )\right]
\,,
\end{align}
where $\vek{s} \equiv \vek{\cal S} / {\cal S}$,
$\vek{k} \cdot \vek{s} = \cos\alpha$ 
and the coefficient $B$ is given here
in terms of the \emph{non-reduced} variables, 
used by Damour and Sch\"afer in Ref.~\cite{DS88} 
and in Appendix~\ref{appendix_LRL_vector}, by
\begin{align}
\label{eq:coefficient_B_DS88}
B &
= \frac{ n \lambda_\text{so} {\cal S} }{ c^2 M (1 - e^2)^{3/2} } 
\,.
\end{align}
The dimensionless mass dependent coupling constant
$\lambda_\text{so}$ is related to $\chi_\text{so}$ by 
$\lambda_\text{so} = \chi_\text{so} / \eta$
(see Appendix~\ref{appendix_LRL_vector} for details).
To check the above agreement between our result and 
the one of Damour and Sch\"afer, 
we have used 
Eqs.~\eqref{eq:theta_in_terms_of_SLJ} and 
Eq.~\eqref{eq:cos_i_para}. 
The two expressions for the coefficient $B$, 
namely Eqs.~\eqref{eq:coefficient_B_CK} and 
\eqref{eq:coefficient_B_DS88}, are identical.

Let us now briefly consider the nonspinning 3PN accurate
contributions to $k$.
These contributions, enter only via $\Pi$,
orbital averaged analogue to Eq.~\eqref{eq:so_contr_to_the_aver_Pi},
our computation,
and result in $k_\text{3PN} \equiv \tilde{k}_\text{3PN}$.
However, it should be noted that the expression
for $k_\text{3PN}$, as given by Eq.~\eqref{eq:just_parameter_k_3PN},
is not yet ready for application.
This is so as this expression for $k_\text{3PN}$
depends on the conserved \emph{theoretical} quantities $E$ and $L$
rather than on \emph{observable} quantities like
mean motion $n$ and proper-time eccentricity $e_T$.
However, we have checked that up to 2PN order our expression
for $k_\text{3PN}$ is in agreement with the one
of Damour and Sch\"afer, obtained in Ref.~\cite{DS88}.

Therefore, the 3PN accurate observable 
periastron advance parameter $k$ becomes
\begin{align}
\label{eq:k_result}
k = k_\text{3PN} + k_\text{so}
\,,
\end{align}
where $k_\text{3PN}$ is the \emph{orbital} 3PN contribution,
corresponding to the nonspinning case,
given by the 3PN accurate contribution to
Eq.~\eqref{eq:parameter_k_3PN_with_SO},
namely Eq.~\eqref{eq:just_parameter_k_3PN}.
The spin-orbit contribution $k_\text{so}$ is given by 
Eqs.~\eqref{eq:k_obs_CK} and \eqref{eq:coefficient_B_CK}
or equivalently by
Eqs.~\eqref{eq:k_obs_DS88} and \eqref{eq:coefficient_B_DS88}. 
We recapitulate that 
$k_\text{3PN} \equiv \tilde{k}_\text{3PN}$ and 
$k_\text{so} \neq \tilde{k}_\text{so}$, where 
$\tilde{k}_\text{3PN}$ and $\tilde{k}_\text{so}$ are
the corresponding contributions to $\tilde{k}$,
as given by Eqs.~\eqref{eq:parameter_k_3PN_with_SO} and
\eqref{eq:parameter_k_3PN_and_k_so}.
This results from the construction of $\tilde{k}$
in Ref.~\cite{KG05spinorbit}
by combining the parameter $\tilde{k}_\text{3PN} = k_\text{3PN}$ 
of the nonspinning case with the internal parameter
$\tilde{k}_\text{so}$, given by Eq.~(4.45e) in Ref.~\cite{KG05spinorbit}, 
of the spin-orbit parametrization
[see Sec.~V in Ref.~\cite{KG05spinorbit} for details].

Finally, let us present the following geometric explanation
why $\tilde{k} \neq k$.
The difference between $\tilde{k}$ and $k$
results from the fact, that the angle $\Pi$ 
--- associated with the parameter $\tilde{k}$ ---
is measured from $\vek{i}$ and in contrast to that,
the angle $\omega$ 
--- associated with the observable parameter $k$ ---
is measured with respect to $\vek{i}_{i}$. 
The difference between $\omega$ and $\Pi$ is denoted by 
$\Delta$ in Fig.~\ref{fig:3fullplanes} 
(see Appendix~\ref{appendix_binary_geometry} for details).
In addition, we want to point out, 
that the last rotation ${\cal R}[\omega(t),\vek{k}(t)]$
in Eq.~\eqref{eq:def_abk_from_pqN_via_DS88}
can be decomposed into two parts
\begin{align}
{\cal R}[\omega(t),\vek{k}(t)]
&=
{\cal R}[\Pi(t),\vek{k}(t)] \;
{\cal R}[\Delta(t),\vek{k}(t)] 
\,,
\end{align}
and so it is obvious that $\omega(t) = \Delta(t) + \Pi(t)$
and correspondingly we have $d\omega/dt = d\Delta/dt + d\Pi/dt$.
The contribution of $d \Delta / dt$ causes the difference between
$k_\text{so}$ and $\tilde{k}_\text{so}$.

%%%%%%%%%%%%%%%%%%%%%%%%%%%%%%%%%%%%%%%%%%%%%%%%%%%%%%%%%%%%%%%%%%%%%%%%%%%%%

\section{Conclusions}
\label{SecConclusions}

We have presented a new method to compute the PN accurate 
observable dimensionless relativistic 
periastron advance parameter $k$ of a spinning compact binary
using our Keplerian-type parametric solution
to the PN accurate conservative dynamics
of spinning compact binaries 
moving in eccentric orbits \cite{KG05spinorbit}.
The expression for $k$ is applicable,
like our parametric solution,
only for the following two distinct cases, namely, 
case (i), the equal-mass case and 
case (ii), the single-spin case.
Our result is consistent with a more general formula for $k$,
derived by Damour and Sch\"afer,
using a different procedure \cite{DS88}.
The computation also helped us to clarify 
certain intriguing features 
that we observed in our parametrization.

%%%%%%%%%%%%%%%%%%%%%%%%%%%%%%%%%%%%%%%%%%%%%%%%%%%%%%%%%%%%%%%%%%%%%%%%%%%%%

\begin{acknowledgments}

It is our pleasure to thank
Gerhard Sch\"afer for helpful discussions and
encouragements. We gratefully acknowledge the financial support of
the Deutsche Forschungsgemeinschaft (DFG) through SFB/TR7
``Gravitationswellenastronomie''.

The algebraic computations, appearing in this paper, were performed
using \textsc{Maple} and \textsc{Mathematica}.

\end{acknowledgments}

%%%%%%%%%%%%%%%%%%%%%%%%%%%%%%%%%%%%%%%%%%%%%%%%%%%%%%%%%%%%%%%%%%%%%%%%%%%%%
%%%%%%%%%%%%%%%%%%%%%%%%%%%%%%%%%%%%%%%%%%%%%%%%%%%%%%%%%%%%%%%%%%%%%%%%%%%%%

\appendix

\section{Some aspects of the binary geometry and related dynamics}
\label{appendix_binary_geometry}

In this section, we make contact with the variables and angles,
commonly used in the literature \cite{DS88,LBK1995}.
It is obvious that we have three planes in the binary geometry, namely
the orbital plane, the invariable plane and the plane of the sky.
The associated orthonormal triads are 
$(\vek{i},\vek{j},\vek{k})$, 
$(\vek{e}_{X},\vek{e}_{Y},\vek{e}_{Z})$, and 
$(\vek{p},\vek{q},\vek{N})$.
Therefore, we may define three lines of nodes and 
associated inclination angles in the following way.
We infer from Fig.~\ref{fig:3fullplanes} that 
the plane of the sky meets the invariable plane along a line
of nodes, characterized by $\vek{p} = \vek{e}_{X}$, as we choose 
the line-of-sight unit vector $\vek{N}$ to lie in the
$\vek{e}_{Y}$-$\vek{e}_{Z}$ plane.
The associated constant angle $i_{0}$ between
$\vek{N}$ and $\vek{J} = J \vek{e}_Z$ may be treated 
as the inclination angle, related to the above line of nodes.
In the second case, we consider the line of nodes $\vek{i}$,
created by the intersection of the orbital plane with the invariable plane.
The corresponding inclination angle, namely the angle between
the orbital angular momentum unit vector $\vek{k} = \vek{L}/L$ 
and $\vek{J} = J \vek{e}_Z$, is denoted by $\Theta$.
Finally, we consider the intersection of the orbital plane
with the plane of the sky.
We associate $\vek{i}_{i}$ with this line of nodes and the related
inclination angle $i$, the angle between
$\vek{k}$ and the line of sight unit vector $\vek{N}$,
is the usually defined orbital inclination $i$
and hence the use of the subscript $i$ in $\vek{i}_{i}$.

It is easy to figure out that the unit vectors,
associated with the last two
lines of nodes, $\vek{i}$ and $\vek{i}_{i}$ are defined by
\begin{subequations}
\begin{align}
\label{eq:def_vector_iTheta}
\vek{i} & 
\equiv \frac{\vek{e}_Z \times \vek{k}}{|\vek{e}_Z \times \vek{k}|}
\,,
\\
\label{eq:def_vector_ii}
\vek{i}_{i} &
\equiv \frac{\vek{N} \times \vek{k}}{|\vek{N} \times \vek{k}|}
\,.
\end{align}
\end{subequations}
Using $\vek{k}$, expressed in terms of
$(\vek{e}_{X},\vek{e}_{Y},\vek{e}_{Z})$ and
$(\vek{p},\vek{q},\vek{N})$, given by
\begin{align}
\label{eq:vector_k_in_xyz_pqN}
\vek{k} &
= S_\Theta \sin\Upsilon \vek{e}_{X}
- S_\Theta \cos\Upsilon \vek{e}_{Y}
+ C_\Theta \vek{e}_{Z}
\nonumber \\
&
= S_\Theta \sin\Upsilon \vek{p}
+ (-S_{i_{0}} C_\Theta - C_{i_{0}} S_\Theta \cos\Upsilon ) \vek{q}
\nonumber \\
&\quad
+ ( C_{i_{0}} C_\Theta - S_{i_{0}} S_\Theta \cos\Upsilon ) \vek{N}
\,,
\end{align}
we compute the following expressions 
for $\vek{i}$ and $\vek{i}_{i}$ as
\begin{subequations}
\begin{align}
\vek{i} 
&= \cos\Upsilon \vek{e}_{X} + \sin\Upsilon \vek{e}_{Y}
\nonumber \\
&= \cos\Upsilon \vek{p} + C_{i_{0}} \sin\Upsilon \vek{q} + S_{i_{0}}
\sin\Upsilon \vek{N}   
\,,
\\
\label{eq:def_vector_ii_in_pqN}
\vek{i}_{i} &
= \frac{1}{\sin i}
\left[
( S_{i_{0}} C_\Theta + C_{i_{0}} S_\Theta \cos\Upsilon)\vek{p}
+ S_\Theta \sin\Upsilon \vek{q}
\right]
\,,
\end{align}
\end{subequations}
where $\sin i = | \vek{N} \times \vek{k} |$.
The time evolution for the (observational) orbital inclination 
$i$ is simply given by
\begin{subequations}
\label{eq:time_evolution_of_i}
\begin{align}
\label{eq:cos_i_para}
\cos i
&= \vek{N} \cdot \vek{k}
= C_{i_{0}} C_\Theta - S_{i_{0}} S_\Theta \cos\Upsilon 
\,,
\\
\label{eq:sin_i_parametrized}
\sin i
&= | \vek{N} \times \vek{k} |
\nonumber \\
&= \left[ ( S_{i_{0}} C_\Theta + C_{i_{0}} S_\Theta \cos\Upsilon )^2
+ S_\Theta^2 \sin^2 \Upsilon \right]^{1/2} 
\nonumber \\
&= \left[ 1 - (C_{i_{0}} C_\Theta - S_{i_{0}} S_\Theta \cos\Upsilon )^2
\right]^{1/2} 
\,.
\end{align}
\end{subequations}
The equation for $\cos i$ is in agreement
with Eq.~(6) in Ref.~\cite{LBK1995}.
We may also obtain a differential equation for the 
time evolution of $i$, as given in Ref.~\cite{DS88}. 
To get that, we differentiate $\cos i = \vek{N} \cdot \vek{k}$
with respect to time and make use of
$\dot{\vek{k}} = \dot{\vek{L}}/L$ with 
$\dot{\vek{L}} = \vek{S}_\text{eff} \times \vek{L} / (c^2 r^3)$,
$\sin i = | \vek{N} \times \vek{k} |$
and Eq.~\eqref{eq:def_vector_ii}.
In this way we deduce the following differential equation for 
the orbital inclination $i$: 
\begin{align}
\frac{d i}{dt} &
= \frac{1}{c^2 r^3} \vek{S}_\text{eff} \cdot \frac{\vek{N} \times
\vek{k}}{\sin i}
= \frac{1}{c^2 r^3} \vek{S}_\text{eff} \cdot 
\vek{i}_{i}
%\frac{\vek{N} \times \vek{k}}{|\vek{N} \times \vek{k}|}
\,,
\end{align}
which agrees with Eq.~(5.15) of Ref.~\cite{DS88}, whereas
we note that the quantity characterizing the
\emph{ascending node} in Ref.~\cite{DS88},
namely $\vek{i}$, is identical to our $\vek{i}_{i}$.

Following Ref.~\cite{DS88}, let $\Omega$ 
be the longitude of the line of nodes $\vek{i}_{i}$,
as measured from $\vek{p}$ in the $\vek{p}$-$\vek{q}$ plane 
(plane of the sky), 
which implies
\begin{align} 
\vek{i}_{i} & = \cos\Omega \, \vek{p} + \sin\Omega \, \vek{q}
\,.
\end{align}
The time evolution for $\Omega$  
is defined by either of the following two equations
\begin{subequations}
\begin{align}
\cos\Omega
&=\vek{p} \cdot \vek{i}_{i}
= \frac{1}{\sin i}
( S_{i_{0}} C_\Theta + C_{i_{0}} S_\Theta \cos\Upsilon )
\,,
\\
\sin\Omega
&= | \vek{p} \times \vek{i}_{i} |
= \frac{1}{\sin i} S_\Theta \sin\Upsilon
\,.
\end{align}
\end{subequations}

Finally, we note that it is also possible to obtain the time evolution
for the longitude of periastron $\omega$, measured from
$\vek{i}_{i}$ in the orbital plane, as defined in Ref.~\cite{DS88}.
To do that, it is advantageous to define a periastron-based orthonormal triad
$(\vek{a},\vek{b},\vek{k}) = (\vek{a},\vek{k} \times \vek{a},\vek{k})$.
This is easily obtained by a rotation of the
$(\vek{i},\vek{j},\vek{k})$ frame around the orbital
angular momentum unit vector $\vek{k}$, by an angle $\Pi$.
This gives us 
\begin{align}
\label{eq:definifition_abk}
\begin{pmatrix}
\vek{a}\\
\vek{b}\\
\vek{k}
\end{pmatrix}
&=
\begin{pmatrix}
\cos\Pi&\sin\Pi&0\\
-\sin\Pi&\cos\Pi&0\\
0&0&1
\end{pmatrix}
%-----------------------
\begin{pmatrix}
\vek{i}\\
\vek{j}\\
\vek{k}
\end{pmatrix}
\,,
\end{align}
where $\Pi$ is the angle between $\vek{i}$ and the
(instantaneous) periastron unit vector $\vek{a}$.
The time evolution of $\omega$ is given by 
either of the following relations
\begin{subequations}
\label{eq:cos_omega_sin_omega}
\begin{align}
\cos\omega
&= \vek{i}_{i} \cdot \vek{a}
%\nonumber
%\\
%&
= \frac{1}{\sin i}
\left[
- S_{i_{0}} \sin\Upsilon \sin\Pi
\right.
\nonumber \\
&\quad\left.
+ (C_{i_{0}} S_\Theta + S_{i_{0}} C_\Theta \cos\Upsilon)
\cos\Pi
\right]
\,,
\\
\label{eq:sin_omega_in_appendix}
\sin\omega
&= |\vek{i}_{i} \times \vek{a}|
%\nonumber
%\\
%&
= \frac{1}{\sin i}
\left[( C_{i_{0}} S_\Theta + S_{i_{0}} C_\Theta  \cos\Upsilon)
\sin\Pi
\right.
\nonumber \\
&\quad\left.+ S_{i_{0}} \sin\Upsilon \cos\Pi
\right]
\,.
\end{align}
\end{subequations}
The parametric solution to $\Pi$, which also gives its time evolution, 
is given by our parametrization via Eq.~\eqref{eq:ansatz_for_varphi}.
In addition, the parametric solution to $\Pi$ can be also obtained by noting 
that $\vek{a} = \vek{\cal A}/|\vek{\cal A}|$, 
where $\vek{\cal A}$ is the Laplace-Runge-Lenz vector given by
\begin{align}
\vek{\cal A} &= 
\vek{\cal P} \times (\vek{\cal R} \times \vek{\cal P}) 
- \mu^2 G M \frac{\vek{\cal R}}{\cal R}
\nonumber \\
&= \mu^2 G M 
\left[ \vek{p} \times (\vek{r} \times \vek{p}) - \frac{\vek{r}}{r} \right]
\,.
\end{align}
Further details are described in Appendix~\ref{appendix_LRL_vector}.

Eq.~\eqref{eq:sin_omega_in_appendix} for $\sin\omega$ is in agreement 
with Eq.~(6) in Ref.~\cite{LBK1995}. The additional check 
$\sin^2 \omega + \cos^2 \omega =1$ leads to
\begin{align}
\sin i
= \left[ 
(C_{i_{0}} S_\Theta + S_{i_{0}} C_\Theta \cos\Upsilon )^2
+ S_{i_{0}}^2 \sin^2 \Upsilon \right]^{1/2}
\,,
\end{align}
which is equivalent to Eq.~\eqref{eq:sin_i_parametrized}
(using trigonometric relations).

Since $\vek{i}_{i}$, $\vek{i}$, and $\vek{a}$ lie in the
orbital plane, we may define $\Delta$ such that it is the angle between
$\vek{i}_{i}$ and $\vek{i}$.
This implies that $\omega = \Pi + \Delta$ (see Fig.~\ref{fig:3fullplanes})
and the time evolution for $\Delta$ is given by
\begin{subequations}
\begin{align}
\cos\Delta &
= \vek{i}_{i} \cdot \vek{i}
= \frac{1}{\sin i} 
( C_{i_{0}} S_\Theta + S_{i_{0}} C_\Theta \cos\Upsilon )
\,,
\\
\sin\Delta &
= |\vek{i}_{i} \times \vek{i}|
= \frac{1}{\sin i} S_{i_{0}} \sin\Upsilon
\,.
\end{align}
\end{subequations}

Finally, let $\varphi_{\text{DS}}$ be the polar angle of $\vek{r}$,
measured from $\vek{i}_{i}$,
as defined by Damour and Sch\"afer in Ref.~\cite{DS88}.
As we define $\varphi$ from our $\vek{i}$, this leads
to $\varphi_{\text{DS}} = \varphi + \Delta$.
Using these inputs, it is possible to write the relative 
separation vector $\vek{r}$ in $(\vek{a},\vek{b})$
in terms of $\varphi_{\text{DS}}$ and $\omega$
as
\begin{align}
\label{eq:r_from_DS88}
\vek{r} &
= r \cos( \varphi_{\text{DS}} - \omega ) \vek{a}
+ r \sin( \varphi_{\text{DS}} - \omega ) \vek{b}
\,,
\nonumber \\
&
= r \cos v \, \vek{a}
+ r \sin v \, \vek{b}
\,,
\end{align}
This is consistent with
Eqs.~(5.5b), (5.6), and (5.10) in Ref.~\cite{DS88}.
In our angular variables, Eq.~\eqref{eq:r_from_DS88} reads
\begin{align}
\vek{r} &
= r \cos( \varphi - \Pi ) \vek{a}
+ r \sin( \varphi - \Pi) \vek{b}
\,,
\nonumber \\
&
= r \cos v \, \vek{a}
+ r \sin v \, \vek{b}
\,.
\end{align}

%%%%%%%%%%%%%%%%%%%%%%%%%%%%%%%%%%%%%%%%%%%%%%%%%%%%%%%%%%%%%%%%%%
% Alternative way via the Laplace-Runge-Lenz vector

\section{The computation of $\dot{\Upsilon}$ and $\dot{\Pi}$ 
via the Laplace-Runge-Lenz vector}
\label{appendix_LRL_vector}

In this section, we present an elegant way,
starting from the famous Laplace-Runge-Lenz vector,
to obtain the time evolution of the  apsidal motion
without any parametrization. The quantities in this section
are non-reduced for a simple comparism with the work
of Damour and Sch\"afer~\cite{DS88}, and the work of 
Wex and Kopeikin~\cite{WexKopeikin1999}.

We define the following functions of the canonically conjugate
phase space variables $\vek{\cal R}$ and $\vek{\cal P}$
\begin{subequations}
\begin{align}
\vek{\cal L}(\vek{\cal R},\vek{\cal P}) 
& \equiv \vek{\cal R} \times \vek{\cal P}
\,, \\
\vek{\cal A}(\vek{\cal R},\vek{\cal P}) 
& \equiv \vek{\cal P} \times (\vek{\cal R} \times \vek{\cal P}) 
- \mu^2 G M \frac{\vek{\cal R}}{\cal R}
\,.
\end{align}
\end{subequations}
The first equation gives 
the orbital angular momentum vector $\vek{\cal L}$
and the second one the apsidal vector $\vek{\cal A}$,
which is the famous supplementary first integral 
%[``Integral'' here is used in the sense of classical mechanics,
%not in the sense of the inverse of differentiation.]
of the Kepler problem.
The vector $\vek{\cal A}$ is usually named after Runge and Lenz
or Laplace, though he was first discovered by Lagrange.

These vectors evolve according to the fundamental equations of 
Hamiltonian dynamics 
\begin{subequations}
\label{eq:dL_dt_and_dA_dt}
\begin{align}
\frac{d \vek{\cal L}}{d T} &
= \{ \vek{\cal L}, {\cal H}\}
\,, \\
\frac{d \vek{\cal A}}{d T} &
= \{ \vek{\cal A}, {\cal H}\}
\,,
\end{align}
\end{subequations}
where $\{ \ldots,\ldots \}$ denotes the Poisson brackets and 
$\cal H$ is the full Hamiltonian for the relative motion of our 
binary system. 
As $\vek{\cal L}$ and $\vek{\cal A}$ are first integrals of
${\cal H}_{\rm{N}}$, only ${\cal H}_{\rm 1PN} + {\cal H}_{\rm 2PN}
+ {\cal H}_{\rm 3PN} + {\cal H}_{\rm SO}$
contribute to the right-hand sides of Eqs.~\eqref{eq:dL_dt_and_dA_dt}.
The effects of ${\cal H}_{\rm 1PN} + {\cal H}_{\rm 2PN}
+ {\cal H}_{\rm 3PN}$ are already studied in the literature, and so
we just want to consider the contribution of
\begin{align}
{\cal H}_{\rm SO} 
(\vek{\cal R}, \vek{\cal P}, \vek{\cal S}_{1}, \vek{\cal S}_{2}) &
= \frac{G}{c^2 {\cal R}^3 }
\vek{\cal L} \cdot \vek{\cal S}_\text{eff}
\,,
\end{align}
where the effective spin $\vek{\cal S}_\text{eff}$ is given by
\begin{align}
\vek{\cal S}_\text{eff} &
\equiv\left( 2 + \frac{3 m_2}{2 m_1} \right) \vek{\cal S}_1
+\left( 2 + \frac{3 m_1}{2 m_2} \right) \vek{\cal S}_2
\,.
\end{align}
We recall that $\vek{\cal R}$, $\vek{\cal P}$, $\vek{\cal S}_1$, and
$\vek{\cal S}_2$ are canonical variables, such that the orbital 
variables commute with the spin variables, e.g.,
see Refs.~\cite{Schaefer_grav_effects_2004,Damour2001}.
Using the basic Poisson brackets 
$\{ {\cal R}^i , {\cal P}_j \} = \delta^i_j$,
$\{ {\cal S}_1^i , {\cal S}_1^j \} = \varepsilon^{ijk} {\cal S}_1^k$,
$\{ {\cal S}_2^i , {\cal S}_2^j \} = \varepsilon^{ijk} {\cal S}_2^k$,
$\{ {\cal R}^i , {\cal R}^j \} 
= \{ {\cal P}_i , {\cal P}_j \}
= \{ {\cal S}_1^i , {\cal S}_2^j \}
= \{ {\cal R}^i , {\cal S}_1^j \}
= \{ {\cal R}^i , {\cal S}_2^j \}
= \{ {\cal P}_i , {\cal S}_1^j \}
= \{ {\cal P}_i , {\cal S}_2^j \}
= 0$, we find
\begin{subequations}
\begin{align}
\label{eq:L_comma_Hso}
\{ \vek{\cal L}, {\cal H}_{\rm SO} \} &
= \frac{G}{c^2 {\cal R}^3 } 
\vek{\cal S}_\text{eff} \times \vek{\cal L}
\,, 
\\
\{ \vek{\cal A}, {\cal H}_{\rm SO} \} &
= \frac{G}{c^2 {\cal R}^3} 
\left[
\vek{\cal S}_\text{eff} \times \vek{\cal A}
- 3 (\vek{\cal L} \cdot \vek{\cal S}_\text{eff})
\vek{\cal L} \times \frac{\vek{n}}{\cal R}
\right]
\,.
\end{align}
\end{subequations}
After orbital averaging these expressions, we can write down 
the equations for the precessional motion, 
induced by the spin-orbit interaction,
in the vectorial form with the help of the 
rotational velocity vector $\vek{\Omega}_{\rm so}$ 
\begin{subequations}
\label{eq:dL_dT_dA_dT_av_with_Omega_SO}
\begin{align}
\label{eq:dL_dT_av_with_Omega_SO}
\left \langle 
\frac{ d \vek{\cal L} }{ d T }
\right \rangle_{T}^{\rm so} &
= \vek{\Omega}_{\rm so} \times \vek{\cal L}
\,, 
\\
\left \langle 
\frac{ d \vek{\cal A} }{ d T }
\right \rangle_{T}^{\rm so} &
= \vek{\Omega}_{\rm so} \times \vek{\cal A}
\,,
\end{align}
\end{subequations}
where
\begin{align}
\label{eq:def_Omega_SO}
\vek{\Omega}_{\rm so} &
= \frac{G}{c^2} 
\left \langle \frac{1}{ {\cal R}^3 } \right \rangle_{T} 
\left[
\vek{\cal S}_\text{eff} 
- 3 \frac{\vek{\cal L} \cdot \vek{\cal S}_\text{eff}}{{\cal L}^2}
\vek{\cal L} 
\right]
\,,
\end{align}
and $\langle \ldots \rangle_T$ denotes the orbital average.

In case of the so-called \emph{simple precession}, 
where the orbital angular momentum $\vek{\cal L}$ 
and the total spin $\vek{\cal S}$
precess about the conserved total angular momentum vector 
$\vek{\cal J} = \vek{\cal L} + \vek{\cal S}$ at the same rate
\cite{KG05spinorbit,Cutler1994,WexKopeikin1999},
the precession of the orbital plane and the longitude of periastron
are best described by the angles $\Upsilon$ and $\Pi$, defined with 
respect to the invariable plane, i.e., the plane perpendicular
to $\vek{\cal J} = {\cal J} \vek{e}_Z$ 
(see Fig.~\ref{fig:3fullplanes} for details):
\begin{subequations}
\label{eq:dL_dt_dA_dt_with_Upsilon_and_Pi}
\begin{align}
\label{eq:dL_dt_with_Upsilon}
\frac{ d \vek{\cal L} }{ d T } &
= \frac{ d \Upsilon }{ d T } \vek{e}_Z \times \vek{\cal L}
\,, \\
\frac{ d \vek{\cal A} }{ d T } &
= \left( \frac{ d \Upsilon }{ d T } \vek{e}_Z 
+ \frac{ d \Pi }{ d T } \vek{k} \right) 
\times \vek{\cal A}
\,,
\end{align}
\end{subequations}
where $\vek{k} = \vek{\cal L} / {\cal L}$ denotes the unit vector in
the direction of the orbital angular momentum vector $\vek{\cal L}$.
The conditions for the \emph{simple precession} are fulfilled
in our restricted cases, namely 
(i), the equal-mass case and 
(ii), the single-spin case.
In these cases the effective spin $\vek{\cal S}_\text{eff}$
and the total spin $\vek{\cal S}$ are related by 
$\vek{\cal S}_\text{eff} = \lambda_\text{so}\vek{\cal S}$,
where the mass dependent coupling constant is given by
\begin{equation}
\label{eq:coupling_constant_lambda_so}
\lambda_\text{so} \equiv
\begin{cases}
2 + 3 / 2 = 7 / 2
& \text{for (i), the equal-mass case}
\,,
\\
2 + 3 m_2 / (2 m_1) 
%2 + \frac{3 m_2}{2 m_1} 
%\; \text{or} \; 
& \text{for (ii),} \, \vek{\cal S}_1 \neq 0 \,, \vek{\cal S}_2 = 0 
\,,
\\
2 + 3 m_1 / (2 m_2)
%2 + \frac{3 m_1}{2 m_2}
& \text{for (ii),} \, \vek{\cal S}_1 = 0 \,, \vek{\cal S}_2 \neq 0 
\,.
\end{cases}
\nonumber
\end{equation}
Note, that $\lambda_\text{so} = \chi_\text{so} / \eta$.
Using $\vek{\cal S}_\text{eff} = \lambda_\text{so}\vek{\cal S}$
in Eq.~\eqref{eq:def_Omega_SO} and comparing
Eqs.~\eqref{eq:dL_dT_dA_dT_av_with_Omega_SO} with 
an averaged version of Eqs.~\eqref{eq:dL_dt_dA_dt_with_Upsilon_and_Pi},
we obtain the precession in terms of the angles $\Upsilon$ and $\Pi$
\begin{subequations}
\label{eq:dotUpsilon_dotPi}
\begin{align}
\left \langle 
\frac{ d \Upsilon }{ d T } 
\right \rangle_{T}^{\rm so} &
= \frac{ G }{ c^2 } 
\left \langle\frac{1}{ {\cal R}^3 } \right \rangle_{T}
\lambda_{\rm so} {\cal J}
\,,
\\
\left \langle 
\frac{ d \Pi }{ d T } 
\right \rangle_{T}^{\rm so} &
= \frac{ G }{ c^2 } 
\left \langle\frac{1}{ {\cal R}^3 } \right \rangle_{T}
\lambda_{\rm so} {\cal L}
\left( 
- 1 - 3 \frac{ \vek{\cal L} \cdot \vek{\cal S} }{ {\cal L}^2 } 
\right)
\,.
\end{align}
\end{subequations}
We note, that in case of $d \Upsilon / d T$ it is even possible 
to obtain a non-averaged equation, if one uses Eq.~\eqref{eq:L_comma_Hso} 
instead of Eq.~\eqref{eq:dL_dT_av_with_Omega_SO} for the comparism with 
Eq.~\eqref{eq:dL_dt_with_Upsilon}.

Our results, given by Eqs.~\eqref{eq:dotUpsilon_dotPi}, are in agreement 
with Eqs.~\eqref{eq:so_contr_to_Pi_Upsilon_Omega_REDUCED} and
with Eqs.~(38) and (42) in Ref.~\cite{WexKopeikin1999}.

%%%%%%%%%%%%%%%%%%%%%%%%%%%%%%%%%%%%%%%%%%%%%%%%%%%%%%%%%%%%%%%%%%

\bibliographystyle{apsrev}

\end{document}